\documentclass[12pt]{article}
\usepackage[utf8]{inputenc}
  \usepackage{graphicx}
  \usepackage{epstopdf} 
  \usepackage{epsfig}
  \usepackage{amsmath}
  \usepackage{amssymb}
  \usepackage{color}
  \usepackage{verbatim}
\usepackage{multirow}
\usepackage[bottom]{footmisc}
\usepackage{cite,url}

\evensidemargin - .5truecm
\newcommand{\be}{\begin{equation}}
\newcommand{\ee}{\end{equation}}
\newcommand{\nn}{\nonumber}
\newcommand{\bea}{\begin{eqnarray}}
\newcommand{\eea}{\end{eqnarray}}
\newcommand{\bfig}{\begin{figure}}
\newcommand{\efig}{\end{figure}}
\newcommand{\bc}{\begin{center}}
\newcommand{\ec}{\end{center}}
\newcommand{\sssty}[1]{\scriptscriptstyle#1}

\def\as{\alpha_s}

\def\sq2{\sqrt{2}}
\newcommand{\smallz}{{\scriptscriptstyle Z}} %
\newcommand{\mz}{m_\smallz}

\newcommand{\smallh}{{\scriptscriptstyle H}}
\newcommand{\mh}{m_\smallh}
\newcommand{\mt}{m_t}

\newcommand{\Ftria}[1]{F^{#1}_\Delta}
\newcommand{\Fbox}[1]{F^{#1}_\Box}
\newcommand{\Gbox}[1]{F^{#1}_2}
\newcommand{\pt}{p_{\scriptscriptstyle T}^2}
\newcommand{\Cnlo}{{\cal C}_{\scriptscriptstyle{\rm NLO}}}

\newcommand{\RG}[1]{{\color{cyan} #1}}

\newenvironment{appendletterA}
 {
  \typeout{ Starting Appendix \thesection }
  \setcounter{section}{0}
  \setcounter{equation}{0}
  
 }{
  \typeout{Appendix done}
 }

\begin{document}\hfill

\begin{flushright}
        \begin{minipage}{3.0cm}
          RM3-TH/16-04
        \end{minipage}        
\end{flushright}

\renewcommand{\thefootnote}{\fnsymbol{footnote}}

\vspace{1cm}
\begin{center}
{\Large \bf On the two-loop virtual QCD corrections to Higgs boson pair 
production in the Standard Model}

\vspace{0.6cm}
{\large Giuseppe~Degrassi$^{\,a,b}$}\footnote{\texttt{degrassi@fis.uniroma3.it}},
{\large Pier Paolo~Giardino$^{\,a,c}$}\footnote{\texttt{pgiardino@bnl.gov}}
{\large and}
{\large Ramona~Gr\"ober$^{\,b}$}\footnote{ \texttt{groeber@roma3.infn.it}},
 \\[7mm]
{\it  (a) Dipartimento di Matematica e Fisica, Universit{\`a} di Roma Tre,
 I-00146 Rome, Italy}\\[2mm]
{\it (b)  INFN, Sezione di Roma Tre, I-00146 Rome, Italy}\\[2mm]
{\it (c) Physics Department, Brookhaven National Laboratory, Upton, New York
11973}

\end{center}
\bigskip
\bigskip
\bigskip
\vspace{1cm}

\centerline{\large\bf Abstract}
\begin{quote}
We compute the next-to-leading order virtual QCD corrections to Higgs pair 
production via gluon fusion. We present analytic results for 
the two-loop contributions to the spin-0 and spin-2 
form factors in the amplitude.  The reducible 
contributions,  given by the double-triangle diagrams, are evaluated exactly 
while the two-loop irreducible diagrams are evaluated
by an asymptotic expansion in heavy top quark mass up to and including terms
of $\mathcal{O}(1/\mt^8)$. Assuming that the  finite top-quark mass effects
are of similar size in the entire range of partonic energies
we  estimate that mass effects can reduce the hadronic
cross section by  at most $10\%$. 
\end{quote}
\thispagestyle{empty}
\vfill
\newpage
\setcounter{page}{1}
\setcounter{footnote}{0}
\renewcommand{\thefootnote}{\arabic{footnote}}
\section{Introduction}
\label{sec1}
After the discovery of the Higgs boson in Run 1 of the
Large Hadron Collider (LHC)~\cite{Aad:2012tfa,Chatrchyan:2012ufa}, 
one of the major targets of Run 2 is the experimental exploration of its properties. In Run 1, the
measured Higgs boson production rate and the extracted values of the
Higgs couplings to fermions and to gauge bosons have 
been  found to be compatible with the predictions of the
Standard Model (SM) within an experimental accuracy of (10\! -- \!20)\%~\cite{ATLAS-CONF-2015-044}.  
On the other hand, the
self-couplings of the Higgs boson, which in the SM
are determined in terms of the mass of the Higgs boson
and the vacuum expectation value of the Higgs field and are thus 
fully predicted, have not been probed yet. They are accessible in multi-Higgs
production processes \cite{Djouadi:1999rca, Muhlleitner:2000jj}
  though a measurement of the quartic Higgs self-coupling lies beyond the 
reach of the LHC \cite{Plehn:2005nk, Binoth:2006ym}. 
Instead, for  the trilinear Higgs self-coupling various studies showed
that it might be accessible at the LHC in Higgs pair production in 
$b\bar{b} \gamma \gamma$~\cite{Baur:2003gp, Baglio:2012np,
Yao:2013ika, Barger:2013jfa, Azatov:2015oxa, Lu:2015jza}, $b\bar{b}
\tau \bar{\tau}$~\cite{Dolan:2012rv, Baglio:2012np}, $b\bar{b}W^+
W^-$~\cite{Papaefstathiou:2012qe} and
$b\bar{b}b\bar{b}$~\cite{deLima:2014dta, Wardrope:2014kya, Behr:2015oqq} 
final states. 

Higgs pair production is not only interesting as a probe of the trilinear Higgs self-coupling, 
but its rate can be significantly modified by new 
physics  effects. For the dominant Higgs pair production mode, gluon fusion, 
this can, for instance, occur due to new loop 
contributions \cite{Dawson:2015oha}, in models with novel 
$hh\bar{t}t$ coupling \cite{Dib:2005re, Grober:2010yv, Contino:2012xk}
 or if the Higgs boson pair
is produced through the decay of a heavy new resonance. The latter two
possibilities can lead to a strong increase of the cross section. First
limits on such scenarios have been given in refs.~\cite{CMS:2014ipa,
  Aad:2014yja, Khachatryan:2015yea, Aad:2015uka, Aad:2015xja}.

A precise prediction of the gluon fusion Higgs-pair production channel 
is essential to constrain new physics or to determine the Higgs 
self-coupling. The gluon fusion process  is 
mediated by heavy fermions via diagrams with box and triangle topologies 
and is hence loop-induced already at the leading order (LO).
In the  ``triangle'' contribution 
a single Higgs boson splits via an $s$-channel exchange into
two Higgs bosons, thus it contains the trilinear Higgs self-coupling.
The ``box'' contribution plays the role of an irreducible background,
 as it does not incorporate the trilinear Higgs self-coupling.
 
In the SM, the LO cross section is fully known since
the late eighties~\cite{Glover:1987nx}. However, similarly to what
happens in single Higgs production, one expects the LO contribution to
be subject to large radiative corrections. A computation of a $2 \to
2$ process at higher orders is extremely challenging. The next-to-leading 
order (NLO) ``triangle''
contribution can be borrowed from the production of a single Higgs
boson ~\cite{Spira:1995rr, Harlander:2005rq, Anastasiou:2006hc,
  Aglietti:2006tp}, whereas a full computation of the NLO ``box'' form
factors is at the moment not available and technically much more
difficult.  Higher order corrections to Higgs pair production are,
however, available in the effective theory with infinite top mass, $\mt$, or,
equivalently, in the limit of vanishing external momentum, at NLO
 \cite{Dawson:1998py} and more recently also at
next-to-next-to-leading order (NNLO)
\cite{deFlorian:2013jea,Grigo:2014jma}.\footnote{For beyond the SM
  extensions, NLO QCD corrections in the limit for vanishing external
  momenta are available for the SM with additional dimension six
  operators~\cite{Grober:2015cwa}, for an additional scalar
  singlet~\cite{Dawson:2015haa}, for the two-Higgs doublet
  model~\cite{Hespel:2014sla}, for composite Higgs models~\cite{Grober:2016wmf},
  for the MSSM \cite{Dawson:1998py, Agostini:2016vze} and
  NMSSM~\cite{Agostini:2016vze}.} 
   Soft gluon resummation at
next-to-next-to-leading-logarithmic (NNLL) accuracy has been performed
in refs.~\cite{Shao:2013bz, deFlorian:2015moa}.  Whereas the
approximation of small external momenta was shown to work quite well
for single Higgs production~\cite{Spira:1995rr}, it can be expected to
be less effective for pair production, due to the larger energy scale
that characterizes the latter process. The approximation can, however,
be improved by factoring out the full LO cross section. The  
error due to the infinite top-mass limit for the part related to  the real 
corrections  has been estimated in 
refs.~\cite{Frederix:2014hta, Maltoni:2014eza}  to be 
roughly $-10\%$ by comparing the $\mt \to \infty$ limit  result with the 
numerical  calculation  of the real corrections with full top-mass dependence.
Instead, the uncertainty of the effective-theory result  for the virtual 
corrections has been estimated  in refs.~\cite{Grigo:2013rya} 
by the  inclusion of higher orders in an expansion in small external momenta 
finding a positive shift  with respect to the $\mt \to \infty$ result. 
This leads to an estimate of 
the uncertainties due to mass corrections at NLO, including also the
the real contributions expanded in small external momenta, of $\pm 10\%$,
with a reduction to  $\pm 5\%$ when the NNLO effective theory result is 
included~\cite{Grigo:2015dia} .

In this paper we reexamine the evaluation of the  virtual NLO QCD corrections
in Higgs pair production. We present an exact result for the reducible 
contribution given by the double-triangle diagrams, while the
irreducible diagrams are evaluated  via an asymptotic expansion in the top 
mass.  Our work differs from the similar previous analyses in 
refs. \cite{Grigo:2013rya, Grigo:2015dia} by the fact that we perform
the asymptotic expansion up to and including terms ${\cal O} (1/\mt^8)$ at 
the level of the amplitudes and not of the cross section, allowing us to 
derive simple analytic expressions for the spin-0 and spin-2 form factors in the
amplitudes. The latter could be used in the future as a check of the result, 
in the relevant center-of-mass partonic energy region, when a complete 
calculation  of the virtual corrections will be available. Furthermore,
our expressions can be easily implemented in Monte-Carlo codes that compute
the hadronic cross section in order to achieve a better description of the 
partonic center-of-mass energy region below the the $2\,\mt$ threshold.

In order to quantify the finite top-mass effects in the NLO corrections
to the  hadronic cross section we make two different comparisons: i) We
compare the NLO  cross sections computed using different orders in the
top-mass expansion. ii) We compare the cross section including the 
${\cal O} (1/\mt^8)$ terms with the one computed factorizing the exact LO cross 
section while evaluating the NLO correction factor in the 
$\mt \to \infty$ limit.   

\par
The paper is organized as follows: in section~\ref{sec2} we give
general formulae for the Higgs pair production cross section. In the
next section we discuss different large-mass evaluations  of the LO cross 
section comparing them with exact result. In
section~\ref{sec4} we outline our method of calculation of the NLO
corrections that are presented in the next section where we also
discuss their numerical impact and the estimate of the error due to the
mass effects in the virtual corrections.
Finally, in section~\ref{sec6} we draw our conclusions. 
The paper is completed with an appendix where we present the analytic
result for the expanded NLO form factors up to and including terms of
${\cal O} (1/\mt^8)$.

\section{Double Higgs Production via gluon fusion \label{sec2}}

In this section we summarize some general results on the
Higgs boson pair production via the gluon fusion mechanism
in proton--proton collisions, $pp \to H H$.  
The hadronic cross section for the process 
$p + p \to H+ H+X$ at center-of-mass energy $\sqrt{s}$, can be 
written as:
\be
M_{HH}^2 \frac{d\,\sigma}{d\, M_{HH}^2}  =  
          \sum_{a,b}\int_0^1 dx_1 dx_2 \,\,f_{a}(x_1,\mu_F^2)\,
         f_{b}(x_2,\mu_F^2) 
\int_0^1 dz~ \delta \left(z-\frac{\tau}{x_1 x_2} \right)
M_{HH}^2 \frac{d\,\hat\sigma_{ab}}{d\, M_{HH}^2} \, ,
\label{sigmafull}
\ee
where $M_{HH}^2$ is the invariant mass of the two Higgs system,
$\tau = M_{HH}^2/s$, $\mu_F$ is the factorization scale,
$f_{a}(x,\mu_F^2)$, the parton density of the colliding proton
for the parton of type $a, \,(a = g,q,\bar{q})$ and $\hat\sigma_{ab}$ is the 
cross section for the partonic subprocess $ ab \to H+ H +X$ at the 
center-of-mass  energy  $\hat{s}=x_1 x_2 s$. The partonic cross section can be 
written in terms of the LO cross section $\sigma^{(0)}$ as:
\be
M_{HH}^2 \frac{d\,\hat\sigma_{ab}}{d\, M_{HH}^2}=
\sigma^{(0)}(z \hat{s})\,z \, G_{ab}(z) \, ,
\label{Geq}
\ee
where, up to NLO terms,
\be
G_{a b}(z)  =  G_{a b}^{(0)}(z) 
          + \frac{\alpha_s (\mu_R)}{\pi} \, G_{a b}^{(1)}(z) \, 
\ee
with $\mu_R$ denoting the renormalization scale.
The LO  contribution is given by the gluon-gluon $(gg)$ channel only,
 i.e.
\be
G_{a b}^{(0)}(z)  =  \delta(1-z) \,\delta_{ag}\, \delta_{bg} \, .
\ee
The  amplitude  for $g_a^\mu (p_1) g_b^\nu (p_2) \to H(p_3) H(p_4)$ 
can be written as:
\be
A^{\mu \nu} = \frac{G_\mu}{\sqrt{2}} \frac{\alpha_s (\mu_R)}{2 \pi} 
\delta_{ab}\, T_F\, 
\hat{s}\left[ A_1^{\mu \nu}  \,F_1 +  A_2^{\mu \nu}\,  F_2
\right]
\label{eqamp}
\ee
where $T_F$  is the matrix normalization factor  for  the fundamental  
representation of $SU(N_c)$ ($T_F =1/2$) and the form factors $F_1, \, F_2$ 
are  functions, besides of $\mt^2$, of the partonic Mandelstam variables
\be
\hat{s} = (p_1 + p_2)^2, ~~\hat{t} = (p_1 - p_3)^2,~~
\hat{u} = (p_2 - p_3)^2~.
\label{Mandvar}
\ee
In eq.~(\ref{eqamp}) the orthogonal projectors $A_1$ and $A_2$ 
onto the spin-0 and spin-2 states, respectively, in $n_d = 4- 2\,\epsilon$ 
dimension and normalized to 2 read 
\bea 
 A_1^{\mu\nu} &= & \sqrt{\frac2{n_d-2}} \left[
g^{\mu\nu}-\frac{p_1^{\nu}\,p_2^{\mu}}{\left(p_1\cdot p_2\right)} \right]
\label{A1}\\
 A_2^{\mu\nu} &= & \sqrt{\frac{n_d-2}{2(n_d-3)}} \left\{
\frac{n_d-4}{n_d-2} \left[
g^{\mu\nu}-\frac{p_1^{\nu}\,p_2^{\mu}}{\left(p_1\cdot p_2\right)} \right] 
\right. \nn \\
&+& \left.
g^{\mu\nu}+ \frac{ 
p_3^2\, p_1^{\nu}\, p^{\mu}_2
- 2\left(p_3\cdot p_2\right)p_1^{\nu}\,p_3^{\mu}
- 2\left(p_3\cdot p_1\right)p_3^{\nu}\,p_2^{\mu}
+ 2 \left(p_1\cdot p_2\right) p_3^{\mu}\,p_3^{\nu}}
{p_{\sssty{T}}^2\left(p_1\cdot p_2\right)}   \right\}
\label{A2}
\eea 
with $p_{\sssty{T}}$ the transverse momentum of the Higgs particle  that can
be expressed in terms of the Mandelstam variables as
\be
p_{\sssty{T}}^2 = \frac{\hat{t}\hat{u} - \mh^4}{\hat{s}}~.
\label{eqpt}
\ee
\begin{figure}[t]
\begin{center}\vspace*{-2.1cm}
\includegraphics[width=0.99\textwidth]{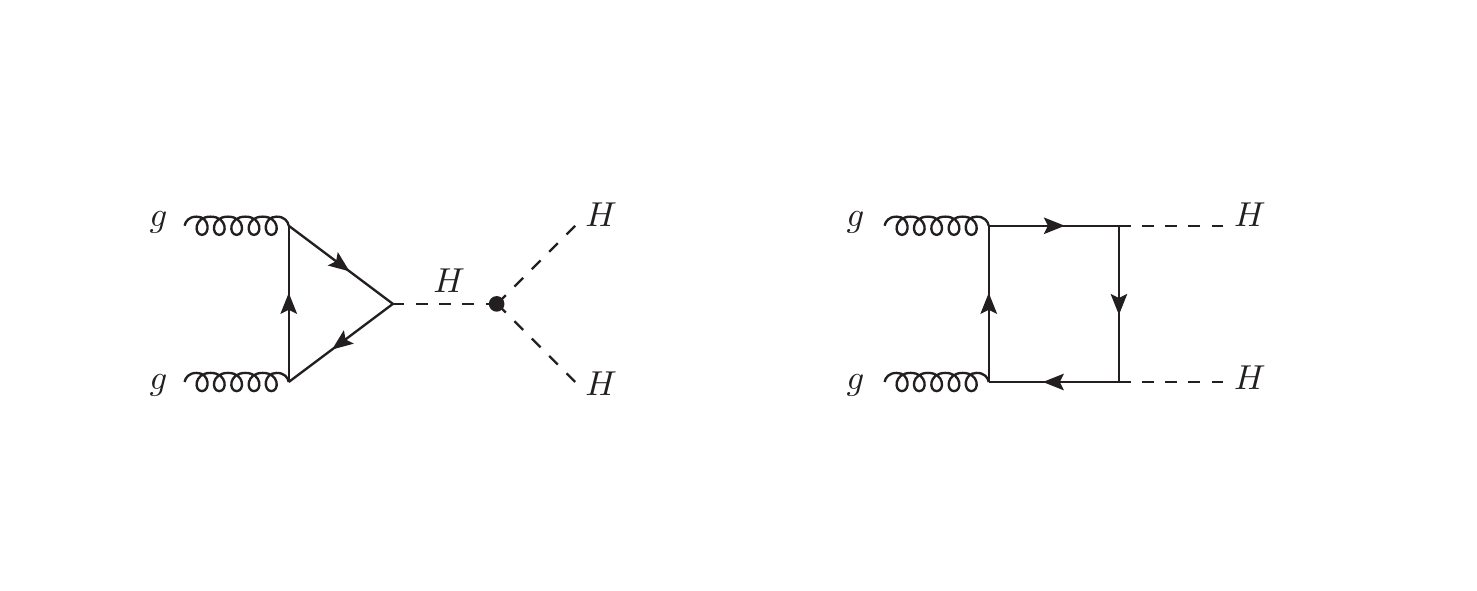}\vspace*{-1.5cm}
\caption{Generic Feynman diagrams for box  and triangle topologies for
Higgs pair production.}
\label{fig1} 
\end{center}
\end{figure}
The spin-2
state receives contributions only from  box topologies (see fig.~\ref{fig1})
 while in the spin-0  case both  box  and triangle diagrams 
contribute such that $F_1$ takes the form
\be
F_1 =   \Ftria{ } \frac{3 \mh^2}{\hat{s} -\mh^2 } 
     +   F_\Box  
\ee
where $\Ftria{} (\Fbox{})$ is the contribution of the triangle (box) diagrams.

The Born cross section is written as 
\bea
\sigma^{(0)}(\hat{s})  =  
\frac{G^2_\mu \alpha_s^2 (\mu_R)}{512\, (2 \, \pi)^3} && \hspace{-0.5cm}
 \int_{\hat{t}_{-}}^{\hat{t}_{+}}   \left\{ 
\left| T_F \, F^{1\ell}_1 (\hat{s}) \right|^2  + 
 \left| T_F\,  \Gbox{1\ell} (\hat{s})\right|^2  \right\} 
\label{ggh}
\eea
with  
$\hat{t}_{\pm} = -\hat{s}/2 (1-2\,\mh^2/\hat{s} \mp \sqrt{1-4\,\mh^2/\hat{s}})$. 
The one-loop form factors $ F_1^{1\ell},\,  F_2^{1\ell}$ are fully known 
analytically \cite{Glover:1987nx,Plehn:1996wb} and their values in the limit
of vanishing external momentum can be obtained via a low energy 
theorem  (LET) calculation~\cite{Ellis:1975ap, Shifman:1979eb, Kniehl:1995tn}
giving $\Ftria{1\ell,LET} =- \Fbox{1\ell,LET} = 4/3, \: \Gbox{1\ell,LET} =0$,
that correspond  to the effective theory $\mt \to \infty$ result.

The NLO terms include, besides the $gg$ channel, also the one-loop
induced processes $gq \rightarrow q H H$ and 
$q \bar{q} \rightarrow g H H$.
The $gg$-channel contribution, involving two-loop
virtual corrections to $g g \rightarrow H H$ and one-loop real
corrections from $ gg \to H H  g$, can be written as 
\bea
G_{g g}^{(1)}(z) & = & \delta(1-z) \left[C_A \, \frac{~\pi^2}3 
 \,+ \beta_0 \, \ln \left( \frac{\mu_R^2}{\mu_F^2} \right) \,+ 
\Cnlo  \right]  \nn \\[1mm]
&+ & 
  P_{gg} (z)\,\ln \left( \frac{\hat{s}}{\mu_F^2}\right) +
    C_A\, \frac4z \,(1-z+z^2)^2 \,{\cal D}_1(z) +  C_A\, {\cal R}_{gg}  \, , 
\label{real}
\eea
where 
\be
\Cnlo = 
\,\frac{\int_{\hat{t}_-}^{\hat{t}_+} d\hat{t}\,\left[ 
\left(T_F\, F_1^{1\ell}\right)^* \,T_F \left( \,F_1^{2\ell} + F_1^{2\Delta} \right)
~+~ \left(T_F\,F_2^{1\ell}\right)^* \,T_F\left(F_2^{2\ell} + F_2^{2\Delta} \right)~
\right]  }{\int_{\hat{t}_-}^{\hat{t}_+} d\hat{t}\, 
\left(\left| T_F \,F_1^{1\ell}\right|^2
+ \left|T_F\,F_2^{1\ell}\right|^2\right)}\, +  \,{\rm h.c.} ~.
\label{nloC}
\ee

In eq.~\eqref{real}, $C_A =N_c$  ($N_c$ being
the number of colors), $\beta_0 = (11\, C_A - 2\, N_f)/6 $ ($N_f$
being the number of active flavors) is the one-loop $\beta$-function
of the strong coupling in the SM, ${\cal R}_{gg}$ is the contribution of the
real corrections, $P_{gg}$ is the LO Altarelli-Parisi
splitting function
\be
P_{gg} (z) ~=~2\,  C_A\,\left[ {\cal D}_0(z) +\frac1z -2 + z(1-z) \right]
\label{Pgg} \, ,
\ee 
and
\be
{\cal D}_i (z) =  \left[ \frac{\ln^i (1-z)}{1-z} \right]_+  \label {Dfun} \, .
\ee

\begin{figure}[t]
\begin{center}\vspace*{-1.0cm}
\includegraphics[width=0.99\textwidth]{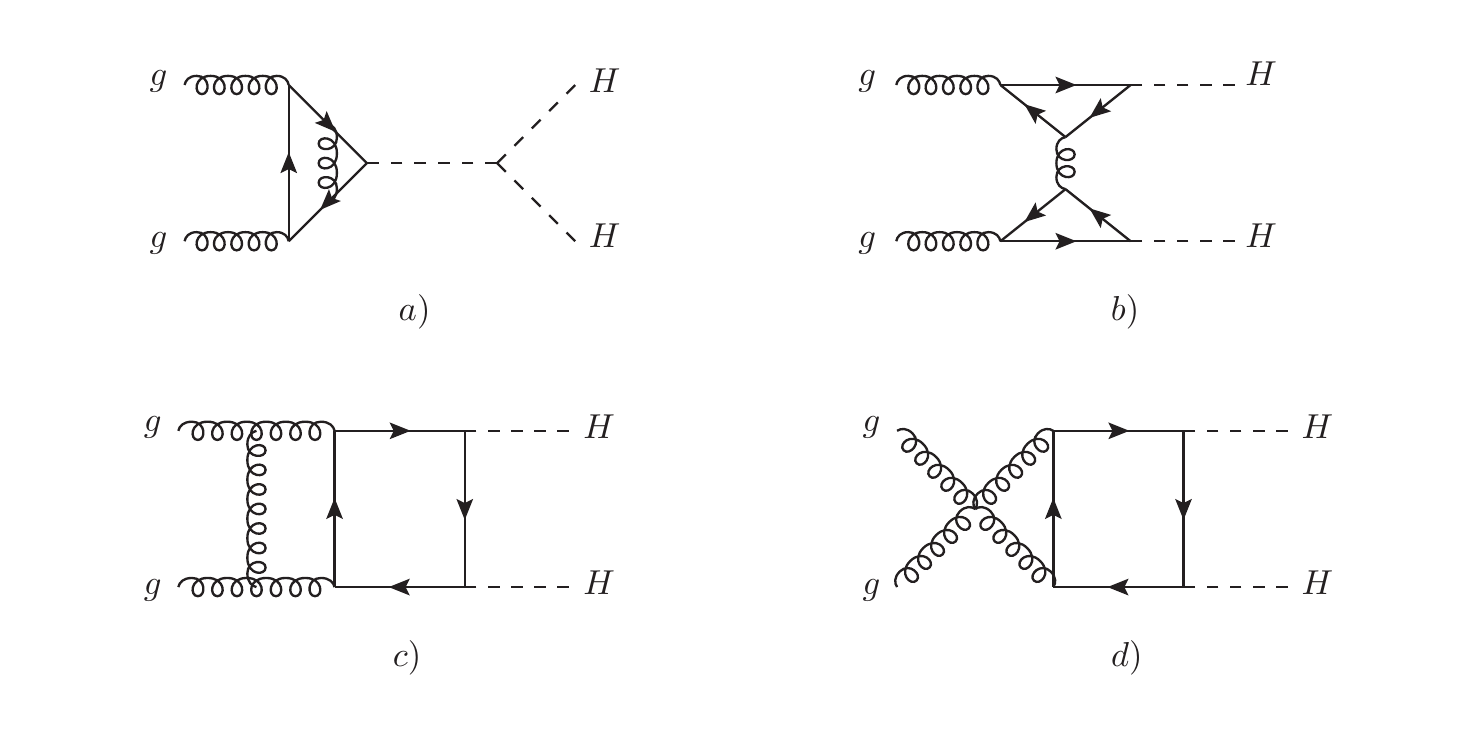}\vspace*{-1cm}
\caption{Sample of Feynman diagrams for the virtual two-loop
corrections to Higgs pair production via gluon fusion.}
\label{fig2} 
\end{center}
\end{figure}

The first line of eq.~\eqref{real} displays the two-loop virtual
contribution regularized by the infrared singular part of the
real-emission cross section.  In eq.~\eqref{nloC} the terms $F_{1}^{2\ell}$ and
$F_{2}^{2\ell}$
contain the contribution of irreducible two-loop diagrams,
(see fig.~\ref{fig2}a,c,d) and  in the limit of vanishing external momenta 
they read
$\Ftria{2\ell} = -\Fbox{2\ell} = - C_F + 5/3 \,C_A, \: \Gbox{2\ell} = 0$ 
\cite{Dawson:1998py}
with    $C_F = (N_c^2-1)/(2\,N_c)$. The term $F_{1}^{2\Delta}\, (F_{2}^{2\Delta})$ 
represents the contribution  of the two-loop
double-triangle diagrams with a $t/u$-channel gluon exchange 
(fig.~\ref{fig2}b)  to the spin-0 (spin-2) part of the amplitude. 
In  the limit of vanishing external momenta the double-triangle diagrams
can be expressed in terms of $F_{\Delta}^{1\ell,LET}$ as
\be
F_1^{2\Delta}\,\rightarrow\, 
\frac12 T_F\,  \left(\Ftria{1\ell, LET}\right)^2~~~
~~~~~~~~ \text{and} ~~~~~~~~~~~F_2^{2\Delta}\,\rightarrow\, 
- \frac12 T_F\,\frac{p_{\scriptscriptstyle T}^2}{2 \hat{t}\hat{u}}(\hat{s}-2\,\mh^2)
 \, \left(\Ftria{1\ell, LET}\right)^2~.
\label{eq:doubletriLET}
\ee

The second line in eq.~\eqref{real} contains the
non-singular contribution from the real gluon emission in the
gluon-fusion process.  The function ${\cal R}_{gg}$ is obtained from
one-loop diagrams where  quarks  circulate in the
loop, and in the limit of vanishing external momenta it becomes 
${\cal  R}_{gg} \to -11 (1-z)^3/(6 z)$. 
The contributions of the $gq \rightarrow q H H$ and $q \bar{q}
\rightarrow g H H$ channels are given by:
\be
G_{q \bar{q}}^{(1)}(z) ~=~   {\cal R}_{q \bar{q}} \, , ~~~~~~~~~~~
G_{q g}^{(1)}(z) ~=~  P_{gq}(z) \left[ \ln(1-z) + 
 \frac12 \ln \left( \frac{\hat{s}}{\mu_F^2}\right) \right] + {\cal R}_{qg} \,,
\label{qqqg}
\ee
where
\be
P_{gq} (z) ~=~  C_F \,\frac{1 + (1-z)^2}z~. 
\ee
The functions ${\cal R}_{q \bar q}$ and ${\cal R}_{q g}$ in
\eqref{qqqg} are obtained from one-loop quark diagrams, and
in the limit of vanishing external momenta become ${\cal R}_{q \bar q}
\to 32 \,(1-z)^3/(27 z)$,~ ${\cal R}_{q g} \to 2\,z/3 - (1-z)^2/z$.

\section{Large mass evaluation of the LO cross section}
\label{sec3}
Even though the one-loop form factors
$ F_1^{1\ell},\,  F_2^{1\ell}$ are fully known analytically
\cite{Glover:1987nx,Plehn:1996wb}, we will give here approximate 
results in order to inspect the validity range of the applied approximations. 
This will later on allow us to apply the same approximations to the NLO cross
section, where the full form factors are yet unknown.

We discuss the large top-mass-expansion evaluation of the LO cross section.
We start by reporting the expressions that 
we obtained via a Taylor expansion for 
$\hat{s}, \hat{t}, \hat{u}, \mh^2 \ll \mt^2$  up to and including 
${\cal O}(1/\mt^8)$ terms
\bea
\Ftria{1\ell} (\hat{s}) &=& \frac43 + \frac7{90} \frac{\hat{s}}{\mt^2} +
               \frac1{126}  \frac{\hat{s}^2}{\mt^4} +
                \frac{13}{12600} \frac{\hat{s}^3}{\mt^6} +
                 \frac{8}{51975} \frac{\hat{s}^4}{\mt^8}  , 
\label{F1tria} \\
\Fbox{1\ell}  (\hat{s}) &=& - \frac43 - \frac7{15} \frac{\mh^2}{\mt^2}
            - \frac{45 \,\mh^4 - 14\, \mh^2 \hat{s} + 6 \hat{s}^2}{315 \,\mt^4}
              + \frac{13}{630} \frac{p_{\sssty{T}}^2\,\hat{s} }{\mt^4} \nn \\
          &&
      -\frac{780 \,\mh^6 -620\, \mh^4 \,\hat{s} + 355\, \mh^2\, \hat{s}^2-
                     16 \, \hat{s}^3}{18900\, \mt^6} -
         \frac{p_{\sssty{T}}^2 ( 11 \,\hat{s}^2 -36 \,\mh^2\, 
             \hat{s})}{1890 \, \mt^6} \nn \\
&& - \frac{ 2400\, \mh^8  - 3480\, \mh^6 \,\hat{s} + 2955\, \mh^4 \,\hat{s}^2
   - 704 \, \mh^2 \,\hat{s}^3 + 120\, \hat{s}^4}{207900 \, \mt^8} \nn \\
  && + \frac{p_{\sssty{T}}^2 \hat{s} (114\, \mh^4 - 85\,  \mh^2\, \hat{s} +
         16 \, \hat{s}^2 - 8\,  p_{\sssty{T}}^2 \,\hat{s})}{10395 \,\mt^8}, 
\label{F1box}\\ 
\Gbox{1\ell} (\hat{s}) &=&  \frac{p_{\sssty{T}}^2}{\mt^2} \left\{ -\frac{11}{45} -
             \frac{ 62\, \mh^2 - 5 \, \hat{s}}{630 \,\mt^2} -
  \frac{400 \, \mh^4 - 156\, \mh^2 \, \hat{s} + 49 \, \hat{s}^2}{ 12600  \,\mt^4}
  + \frac{103}{18900}
         \frac{ p_{\sssty{T}}^2 \,\hat{s}  }{   \,\mt^4}
\right. \nn \\
&&~~~~  \left. 
- \frac{980 \, \mh^6 - 867\, \mh^4 \, \hat{s} + 469\, \mh^2 \, \hat{s}^2  
- 34  \, \hat{s}^3 }{103950  \,\mt^6} + \frac{p_{\sssty{T}}^2\,\hat{s}( 
 24\, \mh^2 \,  -7  \,\hat{s})   }{4950  \,\mt^6}
\right\}~.
\label{G1box}          
\eea 

\begin{figure}[t]
\begin{center}
\includegraphics[width=0.49\textwidth]{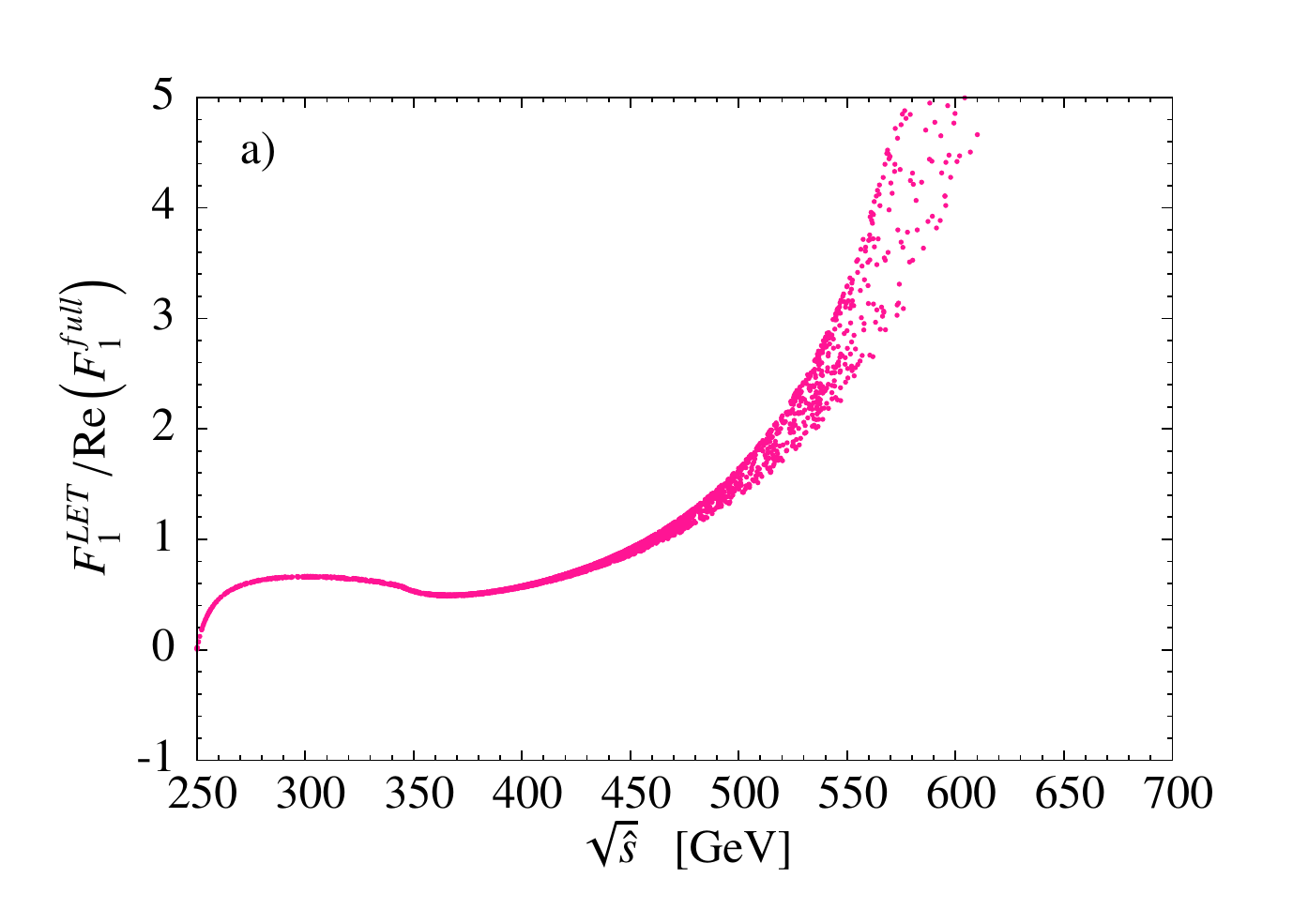}
\includegraphics[width=0.49\textwidth]{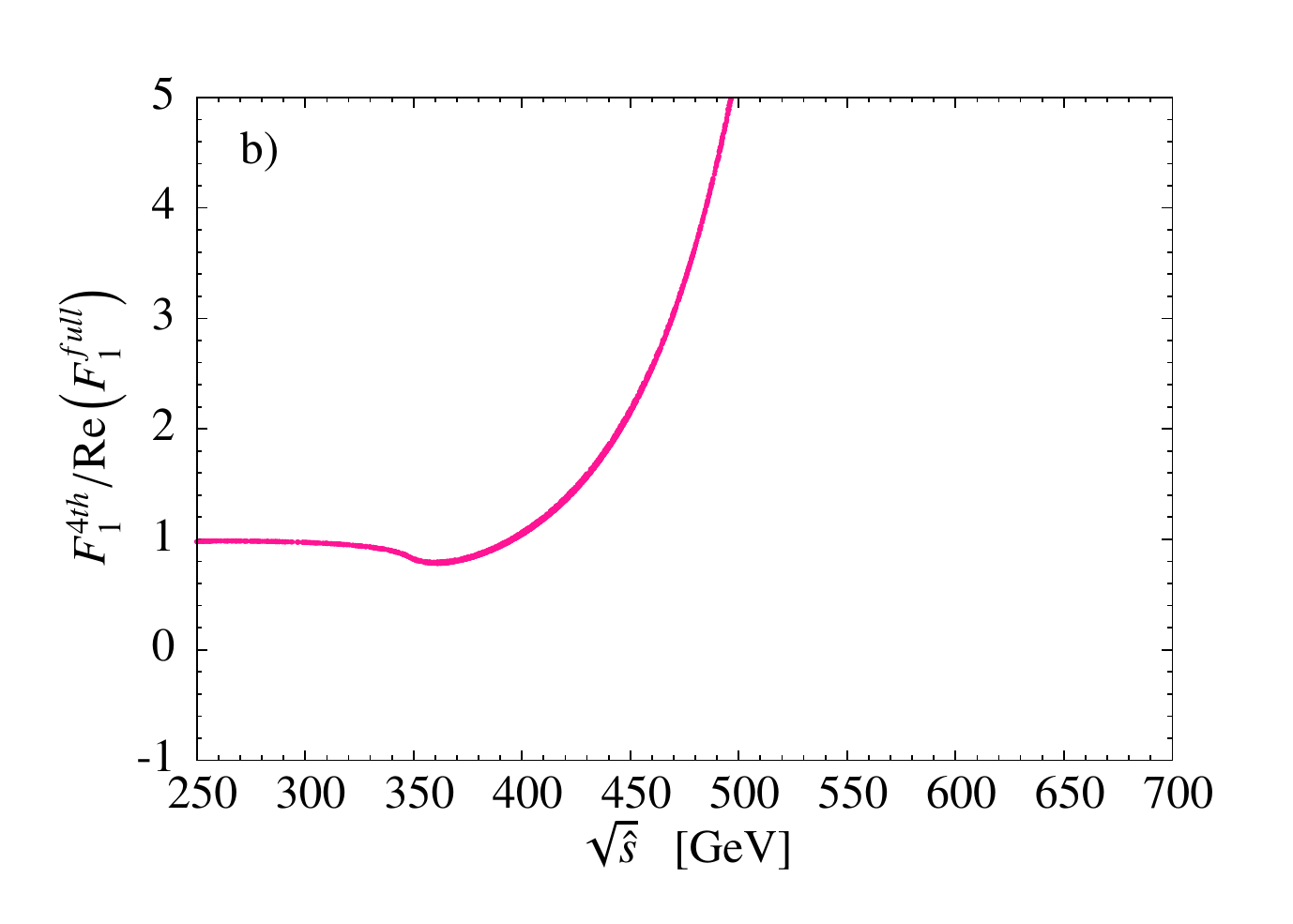}
\caption{ a) LET result 
for $F_1^{1\ell}$ normalized to the real part of the exact $F_1^{1\ell}$ 
form factor.
 b) The sum of first five  terms of the large top-mass expansion  of 
$F_1^{1\ell}$  (eqs.~\eqref{F1tria} and \eqref{F1box}) normalized to the 
real part of  exact $F_1^{1\ell}$ form factor.}

\label{fig3}
\end{center}
\end{figure}

The evaluation of the LO cross section  using for $F_1$ and $F_2$ the values 
obtained via the LET calculation, i.e. the leading term in the large top-mass 
expansion in eqs.~(\ref{F1tria}--\ref{G1box}), 
 gives a  poor approximation of the exact result.   Furthermore, the validity
of this approximation  is quite sensitive to the hadronic center-of-mass
energy and to the choice of the renormalization and factorization scales 
\cite{Dawson:2012mk}. 
This is at variant with the case of single Higgs production where the LET
result gives a quite accurate  estimate of the cross section. Indeed, 
the LET result is expected to be  reliable 
in the region of partonic energies below the $\sqrt{\hat{s}} < 2\, \mt$ 
threshold. In Higgs pair production, also
the region above the $2\, \mt$ threshold contributes significantly to
the hadronic cross section up to $\sqrt{\hat{s}} \sim 600-700$ GeV. In this 
latter region the vanishing external momenta condition  is obviously not 
satisfied
and therefore the  result obtained in this approximation is unreliable.

The inclusion of more terms in a large top-mass
expansion of the form factors does not improve
the evaluation of the LO cross section \cite{Gillioz:2012se, Dawson:2012mk}.
The reason is  easily  understood looking at the plots in Fig.~\ref{fig3}. 
They are obtained evaluating the $F_1$ form factor with 
$p_{\sssty{T}}$ randomly generated but distributed
as for the the integration of the full LO cross section.
The spread in the points for equal $\sqrt{\hat{s}}$ is induced by the 
difference in  the value of $p_{\sssty{T}}$ for fixed  $\sqrt{\hat{s}}$. 
The LET result\footnote{In fig.~\ref{fig3}a the exact cancellation in the LET 
result between the box and the triangle contributions  at the $2\, \mh$ 
threshold is manifest, whereas in the full result the cancellation
between these two contributions is not perfect.} (fig.~\ref{fig3}a) 
approximates relatively well the exact result for $F_1^{1\ell}$ in  the region 
$\sqrt{\hat{s}} \lesssim 2\, \mt$  but it fails in describing  the region 
$\sqrt{\hat{s}} > 2 \, \mt$ when  $\sqrt{\hat{s}} \gtrsim 450$ GeV.
The sum of the first five terms
in the large top-mass expansion of  $F_1^{1\ell}$ 
(fig.~\ref{fig3}b) reproduces quite well the exact results
when $\sqrt{\hat{s}}  \lesssim 400$ GeV while the region 
$\sqrt{\hat{s}} > 400$  GeV is described very badly, worse than in the LET case.
Similar considerations apply to $F_2^{1\ell}$.  

We remark that the evaluation of  $F_1$ and $\Gbox{}$ 
via a large mass expansion has a range of validity up to the 
$ 2 \, \mt$ threshold. Describing the region above this threshold 
via the LET results means to replace the exact  form factors by constant 
values. Instead  using the sum of few terms in the large mass expansion 
means to replace $F_1$ and  $\Gbox{}$ by  a powerlike combination of  
$\hat{s}/\mt^2$ that has a wrong behavior when $\hat{s}$ grows. 
As a consequence,  the partonic cross section in eq.~(\ref{ggh})
grows,  for large values of  the partonic 
center-of-mass energy,   as $\hat{s}$ in the former case, while  as
$\hat{s}^{n+1}/\mt^{2 n}$ in the latter case with $n$  the order of the 
expansion.  Although in both cases the behavior of the partonic cross section
in the region $\sqrt{\hat{s}} > 2\, \mt$ is not described correctly, it is 
evident that in this region the cross section is much better (or less worse)
approximated by its LET value than by including additional terms in the 
large mass expansion. As a further remark, we recall that
the full form factors develop an imaginary part 
above $\sqrt{\hat{s}}> 2\, \mt$ which cannot
be described by an expansion in small external momenta. 
This imaginary part is however smaller than the real part 
up to $\sqrt{\hat{s}}\approx 450 \text{ GeV}$.

\begin{figure}[t]
\begin{center}\vspace*{-1.7cm}
\includegraphics[width=0.75\textwidth]{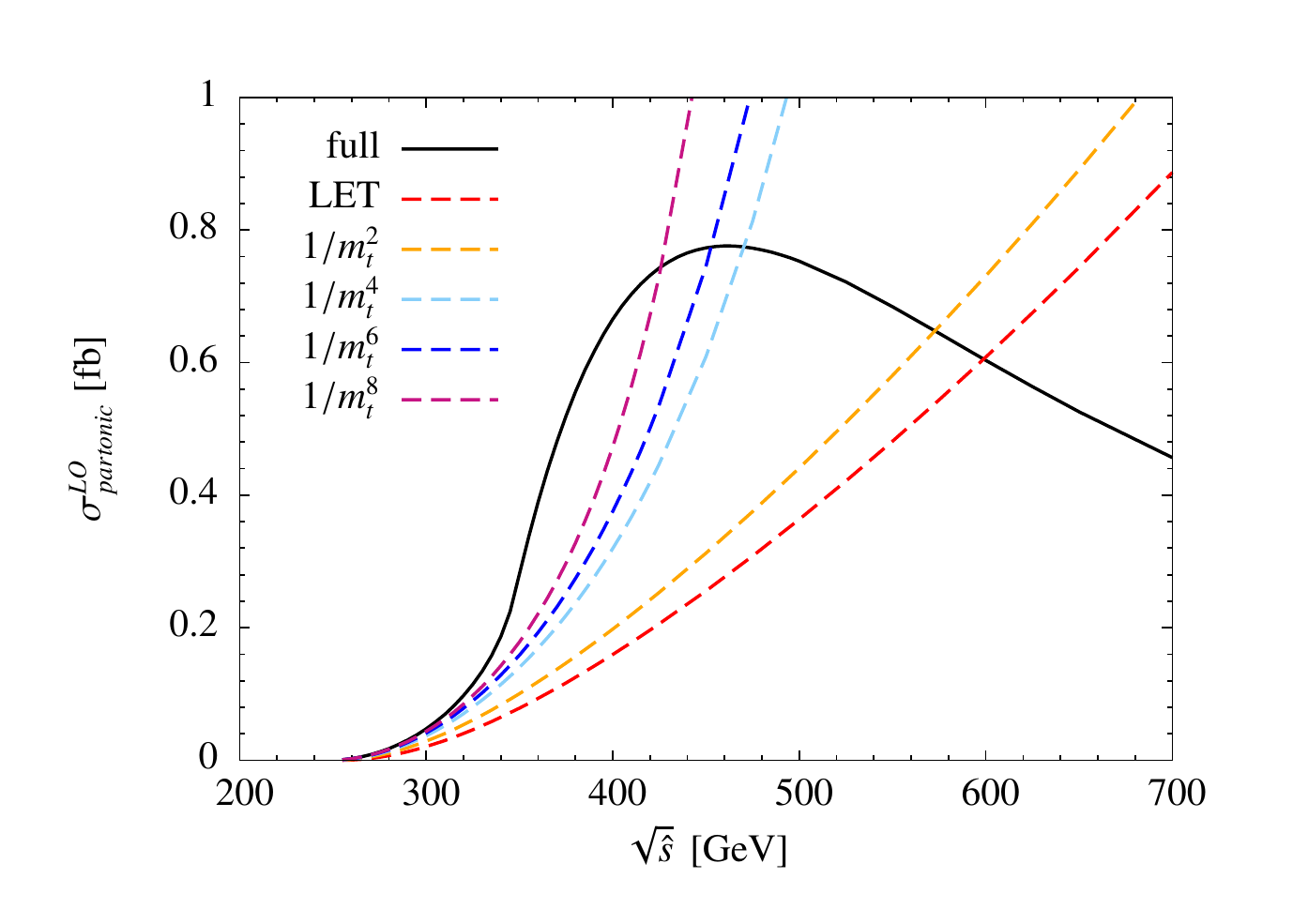}\vspace*{-0.5cm}
\caption{Leading order partonic cross section as a function of the
partonic center-of-mass energy. The solid line corresponds to the exact
result, the dashed ones to the results obtained using different terms 
in the large top-mass expansion. }
\label{fig4} 
\end{center}
\end{figure}

In Fig.~\ref{fig4} we present the partonic cross section as a function of
$\sqrt{\hat{s}}$. The exact cross section (solid black line), 
$\sigma^{(0)}_{ex}$, is compared with the approximated ones 
(dashed colored lines),  $\sigma^{(0)}_{app,n}$, obtained using for the form 
factors the expansions in eqs.~(\ref{F1tria}--\ref{G1box})  to the order $n$. 
The figure tells us  that the validity of an estimate of the hadronic
cross section from eq.~\eqref{sigmafull} based on the use of 
$\sigma^{(0)}_{app,n}$
depends on the relative weights in the hadronic integral of the regions where 
$\sigma^{(0)}_{app,n} < \sigma^{(0)}_{ex}$ vs. 
$\sigma^{(0)}_{app,n} > \sigma^{(0)}_{ex}$
and how these two regions can compensate each other. 
With the increase in the hadronic energy, regions with larger 
$\sqrt{\hat{s}}$ are  going to  contribute more to the hadronic cross section, 
so that   the LET approximation is going to grow in size and therefore 
become either closer to the full cross section or 
overestimating it. For instance for $\sqrt{s}=100 \text{ TeV}$ the LET result 
overestimates the full cross section by a factor $\sim 2.2$.
\begin{table}[h]
  \centering
  \begin{tabular}{|c||c|c|c|c|c|c|}
  \hline
&  $\sqrt{\hat{s}_c}= \infty  $ & 
$\sqrt{\hat{s}_c}= 450  $ & 
$\sqrt{\hat{s}_c}= 400  $
&  $\sqrt{\hat{s}_c} = 350 $   
&  $\sqrt{\hat{s}_c} = 300 $ &  
 $\sqrt{\hat{s}_c}= 0 $\\ \hline \hline
$1/\mt^2$     & 23.18  & 20.12 & 19.80 & 19.49 & 19.25 &  19.17\\ \hline
$1/\mt^4$      & 1703 & 22.63 & 20.96 & 19.90 & 19.32 & 19.17 \\ \hline
$1/\mt^6$   & 4678 & 23.80 & 21.52 & 20.09 & 19.36 & 19.17\\ \hline
$1/\mt^8$ & $7.766\cdot 10^{6}$ & 25.84 & 22.18 & 20.25 & 19.38 &
         19.17 \\ \hline
  \end{tabular}
  \caption{Values in fb of the LO cross section computed 
using  the large mass expansion results of $F_1$ and $F_2$ 
of eqs.~(\ref{F1tria}\,--\,\ref{G1box})
for partonic energies up to $\sqrt{\hat{s}_c} $ (in GeV) 
while  for partonic energies  greater than $ \sqrt{\hat{s}_c} $ 
approximating $F_1$ and $F_2$ with their LET values.}
\label{tab:1}
\end{table}

Figure \ref{fig4}  indicates that an estimate of the LO hadronic cross
section obtained employing the large-mass  expanded results for $F_1$ and $F_2$
in the entire range of partonic energies is not going to be realistic. 
An alternative estimate, based on the use of the maximal approximate information
available and on simplicity, can be obtained by evaluating   
$F_1$ and $F_2$ via a large mass expansion only up to a cut $\sqrt{\hat{s}_c}$
in  the  partonic center-of-mass energy 
while  above $\sqrt{\hat{s}_c}$, where we do not trust any more the expansion,
setting them  to their LET values\footnote{These $F_1$ and $F_2$ functions are 
not continuous at $\sqrt{\hat{s}_c}$.}. This can be considered an improvement
with respect to an evaluation based only on the LET result because  we are
describing better the region $\sqrt{\hat{s}} < 2\,\mt$.  

In Table~\ref{tab:1}  we report  the values of the LO  hadronic cross section 
computed employing different orders in the expansion of  $F_1$ and $F_2$
from eqs.~(\ref{F1tria}\,--\,\ref{G1box}) in the 
region below  $\sqrt{\hat{s}_c}$ while above it the LET values are used.
The values for the cross section are obtained using a modified version of
the code {\tt HPAIR} \cite{HPAIR}, with $\sqrt{s} = 14$ TeV, $\mt=173.2$ GeV, 
$\mh=125$ GeV and employing the parton distribution functions (pdf) 
MSTW08 \cite{Martin:2009iq, Martin:2009bu, Martin:2010db}.
The $\as$ value is taken as the default in the pdf set,
namely $\as^{LO}(\mz) = 0.13939$. 
The renormalization and factorization scales have been set 
to $\mu_R=\mu_F=M_{HH}/2$ as suggested by the 
NNLL threshold expansion performed in ref.~\cite{deFlorian:2015moa}.
The numbers in  the table should be compared with 
the exact LO result\footnote{Note that wherever we use the exact
LO cross section we include always the bottom quark loops.} that, 
including also the bottom contribution, reads
\be
\sigma_{LO}^{full}=23.38 \text{ fb}~. 
\ee
Note that the bottom quark loops contribute with less than 1\%.
One can see from the first column in the table that the use
of the large mass expansion  in the entire range of partonic
energies gives rise to a non convergent result. 
The table also shows that the  if $\sqrt{\hat{s}_c}$ is taken around 
400 GeV the  LO cross section obtained in this way is closer to the exact 
result than  the one that is obtained using  the
LET results (last column of the table).

\section{Outline of calculation \label{sec4}}
An exact analytic evaluation of the two-loop QCD corrections to the
$F_1$ and $F_2$ form factors is presently not available. 
Exact expressions for $F_{1}^{2 \Delta}$ and $F_{2}^{2 \Delta} $ can 
be derived given the structure  of  the double-triangle diagrams 
(fig.~\ref{fig2}b) 
that allows to express the result in terms of products of one-loop 
Passarino-Veltman  functions \cite{Passarino:1978jh}. An exact analytic 
result  for  $\Ftria{2l}$ can be obtained by adapting the corresponding 
calculation in single-Higgs production 
\cite{Spira:1995rr, Harlander:2005rq, Anastasiou:2006hc,Aglietti:2006tp}. 
Instead the exact analytic evaluations of $\Fbox{2\ell}$ and $\Gbox{2\ell}$ 
seem, at the moment, beyond our computational ability. However, it seems 
feasible to obtain an approximate 
evaluation of latter form factors  using the method of asymptotic 
expansions \cite{Smirnov:2002pj,Smirnov:1998vk}. Two different kind of
expansions must  be employed according to the region of partonic energy 
one is considering:
for $\sqrt{\hat{s}} \lesssim 2 \,\mt$ a large mass expansion in the top
mass has to  be performed while in the complementary region 
($\sqrt{\hat{s}} \gtrsim 2 \,\mt$) a large momentum expansion is 
required\footnote{Note that for $\sqrt{\hat{s}} \gtrsim 2 \,\mt$,
different expansions in $\hat{t}/\mt^2$ and $\hat{u}/\mt^2$ need to be 
performed,  depending on whether $\hat{t}$ and $\hat{u}$ are
smaller or larger than $(2\, m_t)^2$.}.
Here we provide a first step in the evaluation of the $\mathcal{O}(\as)$ 
corrections
to $F_1$ and $F_2$ via asymptotic expansions addressing the large mass case.

The large top-mass expansions of the two-loop diagrams contributing to
$\Fbox{2\ell}$ and $\Gbox{2\ell}$  is performed 
using the strategy described in ref.\cite{Degrassi:2010eu} that we 
briefly recall here. The relevant diagrams are generated with
the help of {\tt FeynArts} \cite{Hahn:2000kx}, and contracted with the
projector $A_1^{\mu\nu} \, (A_2^{\mu\nu})$ to extract the $F_1\, (F_2)$ 
contribution. Then  they are  separated  in two
classes: {\it i)} those that can be evaluated via an ordinary Taylor
expansion in powers of $\hat{s}/\mt^2, \, \hat{t}/\mt^2$ and  $\hat{u}/\mt^2$;
{\it ii)} the diagrams that require an asymptotic expansion, i.e.~those that
when Taylor-expanded in the external momenta  exhibit
an infrared (IR) divergent behavior.

Class-{\it i} diagrams require the evaluation of the generic integral 
\be
v(j_1,\dots,j_9,m_1,m_2,m_3) =
\int d^4k_1 \,d^4 k_2 \frac{ (k_1. p_1)^{j_1} (k_1. p_2)^{j_2} (k_1. p_3)^{j_3}
(k_2. p_1)^{j_4} (k_2. p_2)^{j_5} (k_2. p_3)^{j_6} }{
(k_1^2 - m_1^2)^{j_7} (k_2^2 - m_2^2)^{j_8}((k_1+k_2)^2 - m_3^2)^{j_{9}}}
\label{intgen}
\ee
where any exponent $j_1-j_{9}$ is either 0 or a positive integer and 
the propagator masses, $m_1\!-\!m_3$, 
are either $\mt$ or 0.
The integral (\ref{intgen})  can be reduced to 
vacuum integrals, i.e. $v(0,\dots,0,j_7,j_8,j_9,m_1,m_2,m_3)$,  using the 
tensor reduction formula presented in  ref.~\cite{Tarasov:1995jf}.
The two-loop vacuum integrals obtained from the reduction can be evaluated 
using the results of ref.~\cite{Davydychev:1992mt}.

The two-loop diagrams that belong to class {\it ii)} are those containing
either two triple-gluon vertices (fig.~\ref{fig2}c) or one four-gluon
vertex  (fig.~\ref{fig2}d). These diagrams become more and more IR divergent
when the gluon propagators are Taylor-expanded with respect to the 
external momenta.
 The evaluation of the class {\it ii)} diagrams is
obtained by supplementing the Taylor-expanded result by the exact 
computation of their IR divergent contribution.\footnote{The 
second, disconnected, term in part A of fig.~1 of ref.~\cite{Degrassi:2010eu}
gives rise to a vanishing contribution because it contains
scaleless one-loop integrals.} The IR-divergent part of any diagram is 
constructed by  a repeated application of the identity in eq.~(3.1) of
ref.~\cite{Degrassi:2010eu}  controlled by the power counting in the 
IR-divergent terms. The outcome of the procedure is that
the IR-divergent part of any diagram is expressed in terms of products of 
one-loop integrals with numerators that contain terms of the form $(k_{i}\cdot
q_{j})^m \, (k_1 \cdot k_2)^n\,(i =1,2\, , \, j=1,2,3)$ where $m,\, n$ are 
generic powers. Finally, the Passarino-Veltman reduction method is applied to
eliminate the numerators and express the result in terms of the known
one-loop scalar integrals \cite{Passarino:1978jh}. 

\section{Virtual corrections to $gg\to HH$ \label{sec5}}
In this section we give the analytical results for  the double-triangle
form factors $F^{2\Delta}_1$ and $F^{2\Delta}_2$ and the  
two-loop form factors $F_{1}^{2\ell}$ and $F_{2}^{2\ell}$ and
discuss their numerical impact.
\subsection{Analytic results for the two-loop form factors}
We present, for the first time, the exact 
computation of the double-triangle diagrams, i.e. keeping the
full dependence on the quark masses. 
The top contribution to the form factors can be expressed in terms of 
one-loop integrals so that, defining 
\bea
F^{2 \Delta}(x)&=& \frac{8\, m_t^4}{(\mh^2-x)^2} \bigg[1+ \frac{x}{(\mh^2-x)} 
\left(B_0(\mh^2, m_t^2, m_t^2) - B_0(x, m_t^2, m_t^2)\right) \nonumber\\ & &
 +\frac{1}{2} \left(4\, m_t^2-\mh^2+x \right)
  \, C_0(0, x, \mh^2, m_t^2, m_t^2, m_t^2) \bigg]^2~,\label{F2deltahelp}
\eea
we find for $F_{1}^{2\Delta}$ and $F_{2}^{2\Delta}$ in eq.~\eqref{nloC}
\bea
F_1^{2\Delta}&= &    F^{2\Delta}(\hat{t})
+ F^{2\Delta}(\hat{u})~, \label{eq:doubletri1}\\
F_2^{2\Delta}&= & \frac{p_{\scriptscriptstyle T}^2 }{\hat{t}}
F^{2\Delta}(\hat{t})+ \frac{p_{\scriptscriptstyle T}^2 }{\hat{u}}
F^{2\Delta}(\hat{u})~. \label{eq:doubletri2}
\eea
The finite parts of the scalar one-loop integrals appearing 
in eq.~\eqref{F2deltahelp} are given by
\bea
B_0(x, m^2, m^2)&=& 2 +\beta_x \log \frac{\beta_x-1}{\beta_x+1}-\log \frac{m^2}{\mu_R^2}~,\\
C_0(0,x,y,m^2, m^2, m^2)&=&  \frac{1}{2(x-y)}
\left( \log^2\frac{\beta_x+1}{\beta_x-1}-
\log^2\frac{\beta_y+1}{\beta_y-1}\right)~,
\eea
with $\beta_x=\sqrt{1-4\, m^2/x}$ and $\beta_y$ defined in analogy. The bottom
contribution can be obtained thorough the substitution $\mt \to m_b$ in 
eq.~\eqref{F2deltahelp}.
\par
The two-loop form factors $\Ftria{2\ell}, \, \Fbox{2\ell}$ and $\Gbox{2\ell}$
can be written as
\be
 F_i^{2\ell} (\hat{s}) = C_F\, F_{i,C_{F}}^{2\ell}(\hat{s}) + 
          C_A\, F_{i,C_{A}}^{2\ell} (\hat{s})~~~~~~~~~
(i= \Delta,\Box,2)
\label{f2li}
\ee
where $ F_{i,C_{F}}^{2\ell}$ is directly obtained from the two-loop
virtual diagrams and depends upon the renormalized top-mass parameter employed
that we choose to be the on-shell mass. The term $ F_{i,C_{A}}^{2\ell}$ represents
the IR regularized results after subtraction of the IR poles, i.e.
\be
F_{i,C_{A}}^{2\ell} (\hat{s}) = F_{i,C_{A}}^{virt} (\hat{s}) + 
\delta F_{i,C_{A}} (\hat{s})
\label{f2lica}
\ee
where  $F_{i,C_{A}}^{virt}$ is the contribution of the two-loop virtual
diagrams and $ \delta F_{i,C_{A}}$ the counterterm required to make it
finite that reads
\be
\delta F_{i,C_{A}} (\hat{s}) = \frac{1}{2 \epsilon^2} 
F_{i}^{1\ell} (\hat{s},\epsilon)   (\hat{s})^{-\epsilon}
\label{f2lcact}
\ee
where $F_{i}^{1\ell} (\hat{s},\epsilon)$ is the one-loop result including 
the ${\cal O}(\epsilon, \epsilon^2)$ terms.

Employing the method described in sect.~\ref{sec4} we obtained  the large 
top-mass expansion of the two-loop spin-0 and spin-2 form factors 
up to and including terms ${\cal O}(1/\mt^8)$. The results
are presented in appendix~\ref{app:1}. As in the one-loop case the form factors
are expressed in terms of $\hat{s}, \, \pt,\, \mh^2,$ and $\mt^2$.
The computation was performed first using orthogonal projectors in 
$n_d=4 -2\,\epsilon$ dimension (see eqs.~(\ref{A1},\ref{A2})) and then 
using orthogonal projectors in  $n_d=4$ dimension. We  found  that, after the
addition of  the counterterm pieces from  eq.~\eqref{f2lcact}, the two results 
are identical.
We  checked that, once the IR counterterm is chosen as in eq.~\eqref{f2lcact},
$F_{\Delta,C_{A}}^{2\ell}$  reproduces the result for the triangle 
form factor that can be obtained directly adapting the 
known results on single Higgs production (cfr.~eq.~\eqref{fdeltaCA}
 with ref.~\cite{Aglietti:2006tp}). 

\par
Finally, we want to comment on the comparison of our results with the ones of
refs.~\cite{Grigo:2013rya, Grigo:2015dia}. These references deal with the
large top-mass evaluation of the NLO cross section while we concentrated
only on the virtual corrections. We use a different method for
the asymptotic expansion compared to refs.~\cite{Grigo:2013rya, Grigo:2015dia}.
Instead of adding subgraphs and co-subgraphs we 
followed ref.~\cite{Degrassi:2010eu},
where we only add the IR divergent parts, evaluated fully, to the diagrams
that exhibit IR divergencies. We work at the level of amplitudes
while  in ref.~\cite{Grigo:2013rya} 
the total cross section is computed by deriving the imaginary part
of the $gg\to gg$ amplitude and connecting it via the optical theorem
to the total cross section.\footnote{
In \cite{Grigo:2015dia} the virtual corrections have been 
computed also directly from the cut $gg\to HH$ amplitude.}
In refs.~\cite{Grigo:2013rya, Grigo:2015dia}
the phase space integrals are computed analytically which
requires an expansion in $\delta= 1- 4\,\mh^2/\hat{s}$ including
the $s$-channel Higgs 
propagator in the triangle contributions. The result is then expressed as
an expansion both in $\rho=\mh^2/\mt^2$ and $\delta$. 
Our approach is instead 
to compute the  phase space integrals numerically via Monte Carlo methods. 
Performing a Monte Carlo integration of the phase space integrals
 will allow us in future to include expansions
in other regimes or exact results, once available, in a straightforward way.
We remark that the integration over  $\hat{t}$  in eq.~\eqref{nloC}
of the expanded form factors can be easily done analytically as the latter
are given as power series in $\hat{t}$.
A precise numerical comparison with refs.\cite{Grigo:2013rya, Grigo:2015dia} 
cannot be performed, since we did not  compute the real radiation part of 
the $gg$ amplitude. 
 
\subsection{Numerical results}
We discuss now the numerical impact of the corrections we computed.
The numerical results are obtained with a private version of the code 
{\tt HPAIR} \cite{HPAIR} where  we implemented our results. The inputs in
the code are  the same as in table \ref{tab:1}, but using the NLO value for
the strong coupling, $\as^{NLO} (m_Z) = 0.12018$.

\begin{figure}[t]
\begin{center}\vspace*{-1.7cm}
\includegraphics[width=0.78\textwidth]{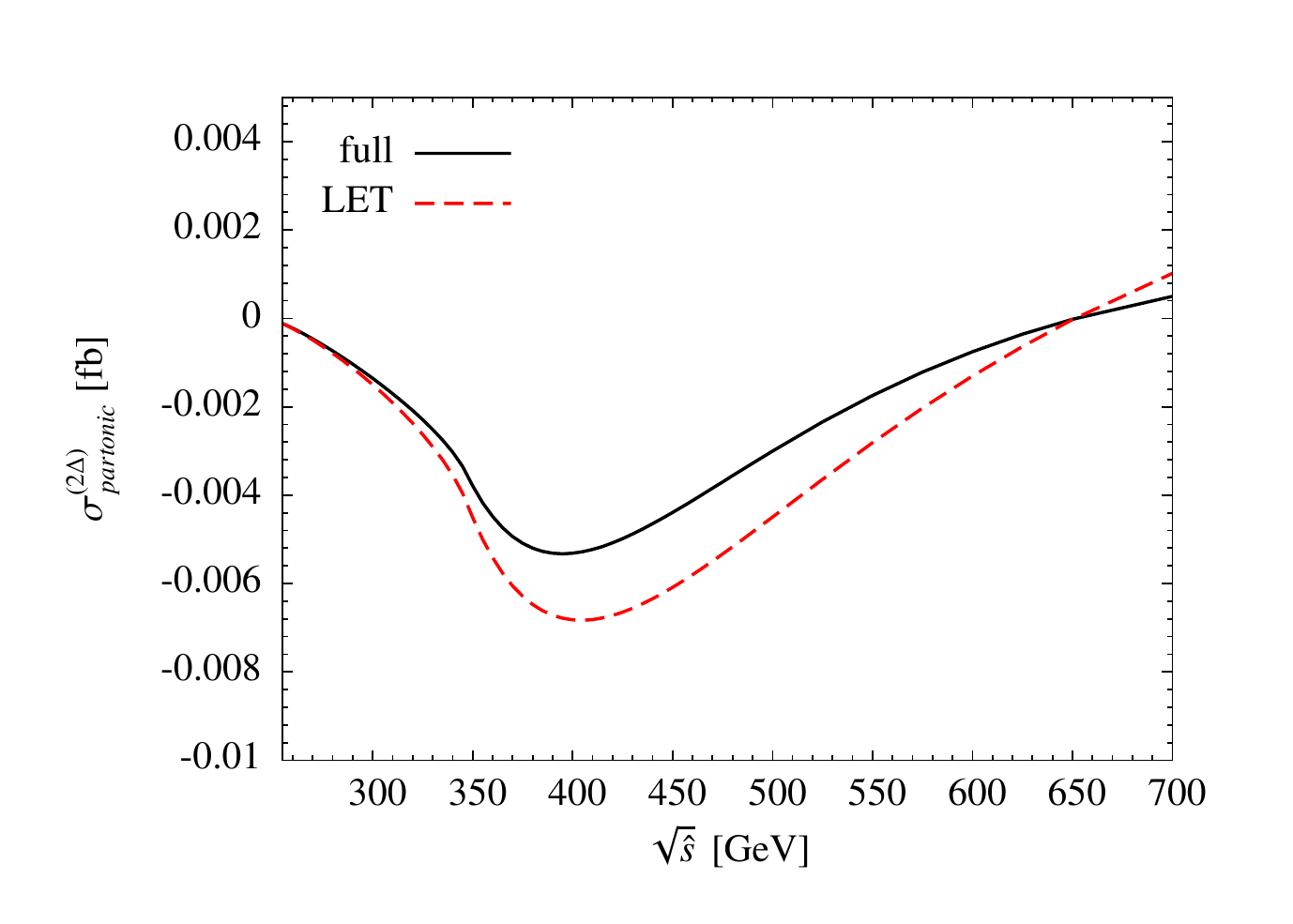}\vspace*{-0.5cm}
\caption{Double-triangle contribution to the  
partonic cross section as a function of the
partonic center-of-mass energy. The solid line represents the exact
result using eqs.~\eqref{eq:doubletri1} and \eqref{eq:doubletri2}
while the dashed one the result obtained in the LET approximation 
using eq.~\eqref{eq:doubletriLET}. }
\label{fig5} 
\end{center}
\end{figure}

We start analyzing the NLO contribution due to the 
double-triangle diagrams, i.e. 
$\sigma^{(2 \Delta)} =\sigma^{(0)} \Cnlo$
with $ F_1^{2\ell}=F_2^{2\ell}=0$ (see eq.~\eqref{nloC}). In 
 order to quantify the impact of the inclusion of the finite 
mass effects  we plot, in fig.~\ref{fig5}, 
$\sigma^{(2 \Delta)}$ computed in two ways: i)  exactly (solid line), i.e. with
all the form factors evaluated in full mass
dependence, namely we use for $F_1^{2\Delta}\,(F_2^{2\Delta})$  
eq.~\eqref{eq:doubletri1} (eq.~\eqref{eq:doubletri2}). ii) Computing
$F_1^{2\Delta}$ and $F_2^{2\Delta}$ using their 
 LET approximation as given in eq.~\eqref{eq:doubletriLET},
while employing the exact expressions for $ F_1^{1\ell}$ and $F_2^{1\ell}$
(dashed line). The figure shows that the inclusion of the finite top mass
effects changes the double-triangle contribution to the partonic cross section
by  $\sim 20-30\%$. We remark that the double-triangle
contribution to the hadronic cross section is actually always  very small. 
Indeed, it amounts  to $\sim -0.18$ 
fb while the  NLO cross section is around 40 fb. 

We turn now to discuss the  contribution in 
$\Cnlo$ due to  $ F_1^{2\ell}$ and $F_2^{2\ell}$
in eq.~\eqref{nloC}. We expect our results for  $ F_1^{2\ell}$ and $F_2^{2\ell}$ 
to be quite accurate for $\sqrt{\hat{s}} \lesssim 400$ GeV, in analogy with 
the LO case as shown in fig.~\ref{fig3}. This allows us to evaluate the 
contribution induced by the mass effects in the virtual part of the 
NLO corrections by 
computing $\sigma^{(0)}\, \Cnlo$ at various order in the large mass expansion.

In table \ref{tab:2} we report the contribution of 
\be
\sigma^{(0)}\, \Cnlo = \frac{G_{\mu}^2\as^2(\mu_R)}{512\, (2\pi)^3} 
\frac{\alpha_s(\mu_R)}{\pi} \int_{t_-}^{t_+}
  2 \,\text{Re} \left(T_F\,F_{1}^{1\ell, full}(T_F\,F\,_1^{2\ell, n})^* +
T_F\,F_{2}^{1\ell, full}(T_F\,F_2^{2\ell, n})^* \right)
\label{SCnlonoi}
\ee
to the hadronic cross section  for few values of an upper cut on the invariant
mass of the two Higgs system\footnote{For the virtual corrections $M_{HH}$ and
$\sqrt{\hat{s}}$ coincide.}, $M_{HH}^c$, and various orders in the expansion.
In eq.~\eqref{SCnlonoi}  $F_{i}^{1\ell, full},\: (i=1,2)$ indicates the exact 
expression of the one-loop form factor \cite{Glover:1987nx,Plehn:1996wb},
 while $F_i^{2\ell, n}$ the expression for the two-loop form factor we derived
(eqs.~(\ref{fdeltaCF}-\ref{f2CA})) to the relevant order $n$. 
Comparing the LET row with the $1/\mt^8$ one, we find that the mass effects 
induce a relative variation with respect to the $\mt \to \infty$ result up 
to $\sim 40 \%$.
\begin{table}[t]
  \centering
  \begin{tabular}{|c|c|c|c|c|}
  \hline
&  $M_{HH}^c = 280  $ 
&  $M_{HH}^c = 300  $ 
&  $M_{HH}^c= 350  $
&  $M_{HH}^c= 400   $ \\
\hline \hline
 LET      & 0.01037  &  0.04103	 &  0.2392 & 0.617   \\ \hline
$1/\mt^2$ &  0.00856 & 0.03454   & 0.1950  & 0.477    \\ \hline
$1/\mt^4$ & 0.01192  & 0.04638   & 0.2784  & 0.775    \\ \hline
$1/\mt^6$ & 0.01335  & 0.05110   &  0.3085 & 0.874   \\ \hline
$1/\mt^8$ & 0.01417  & 0.05445   & 0.3414  & 1.046   \\ \hline
  \end{tabular}
  \caption{Contribution (in fb) of $\sigma^{(0)}\, \Cnlo$ as defined 
in eq.~\eqref{SCnlonoi} 
 to the hadronic cross section for few values of an upper cut on the invariant 
mass of the two Higgs system (in GeV).} 
\label{tab:2}
\end{table}

\begin{table}[t]
  \centering
  \begin{tabular}{|c||c|c|c|c|}
  \hline
&  $M_{HH}^c= 280  $  
&  $M_{HH}^c =300  $ 
&  $M_{HH}^c= 350  $
&  $M_{HH}^c =400  $ \\
 \hline \hline
  LET     & 0.01785 & 0.06534 & 0.3908 & 1.225    \\ \hline
$1/\mt^2$ & 0.01249 & 0.04747 & 0.2880 & 0.870    \\ \hline
$1/\mt^4$ & 0.01296 & 0.05085 & 0.3302 & 1.090   \\ \hline
$1/\mt^6$ & 0.01339 & 0.05221 & 0.3374 & 1.101  \\ \hline
$1/\mt^8$ & 0.01399 & 0.05438 & 0.3587 & 1.222   \\ \hline
  \end{tabular}
  \caption{As Table \ref{tab:2} but with $\sigma^{0}\Cnlo$,  
computed factorizing  the LO cross section (see text).}
 \label{tab:3}
\end{table}

Based on the experience gained in single Higgs production one
expects that the  factorization  of  the exact 
LO cross section can improve the $\mt \to \infty$ determination of the 
hadronic cross section. Applying the same procedure to a large mass expansion 
determination amounts to evaluate $\sigma^{(0)}\, \Cnlo$ employing for 
$\sigma^{(0)}$ the exact LO cross section while evaluating  
$\Cnlo$ at the same order of
approximation both in the numerator and in the denominator.
The contribution to the hadronic cross section of $\sigma^{0} \Cnlo$
computed factorizing  the exact LO cross section  is presented in
table \ref{tab:3}.  Looking at tables \ref{tab:2} and \ref{tab:3} we 
notice that the factorization of the exact LO cross section in the $\mt \to
\infty$ result has the tendency  to overestimate the  NLO cross section 
as approximated by the $1/\mt^8$ rows in the tables. Both tables show 
in the first two columns a good
convergence with respect to the order of the expansion, with the exception
of the $1/\mt^2$ term. As expected, for 
$M_{HH}^c > 2 \, \mt$ the convergence starts to downgrade.

Our analysis cannot say anything about the region of partonic energies
$\sqrt{\hat{s}} \gtrsim 400$ GeV where our results cannot be trusted. 
Concerning the hadronic cross section we can only make a guess
assuming that in both regions, $\sqrt{\hat{s}} \lesssim 400$ GeV and 
$\sqrt{\hat{s}} \gtrsim 400$ GeV, the variation induced by mass effects in
$\sigma^{0}\Cnlo$ will be of a similar size and behaviour so that compensations
between the two energy regions are not going to happen.
 Comparing the first row in table \ref{tab:3} with
the last one in table \ref{tab:2} we find a relative variation $\sim 20\%$.
We notice that the contribution
of  $\sigma^{0}\Cnlo$ to the NLO cross section is about $10\%$ of the total,
that the contributions we did not discuss, i.e. ${\cal R}_{gg}, \:
{\cal R}_{q \bar{q}}$ and ${\cal R}_{qg}$, when evaluated in the limit of
vanishing external momenta contribute to the total NLO cross section by  
$\sim 2 \%$ and that, according to the analysis 
 in refs.~\cite{Frederix:2014hta, Maltoni:2014eza}, the finite mass effects
reduce the size of the real contributions with respect their LET estimate.
Considering a maximal case   we expect that
mass effects are going to reduce the  $\mt \to \infty$  value of the NLO 
cross section by less than $10 \%$. 
This size of variation is indeed found if one compares the NLO cross
section evaluated in the $\mt \to \infty$ limit with the LO term factorized,
$\sigma^{NLO}_{LET} = 40.00$ fb, with the NLO cross section computed as in table
\ref{tab:1} with $\sqrt{\hat{s}_c} = 400$ GeV, the cut in the partonic 
energy that at LO gives a result close to the exact LO value. 
The latter cross section, that is computed evaluating
$F_i^{2\ell}$ in the region below $\sqrt{\hat{s}_c}$ using the $1/\mt^8$ order 
in the  expansion  while above $\sqrt{\hat{s}_c}$ employing the LET values, 
amounts to $\sigma^{NLO}_{\hat{s}_c} = 37.86$ fb.   

Finally we comment on larger hadronic center-of-mass energies. At LO,
the LET result, e.g. at $\sqrt{s}=100$ TeV, approximates the true one worse
than at $\sqrt{s}=14$ TeV.  Even though  a large
center-of-mass energy  gives a stronger weight to the region where
the approximation of large top mass is not valid, we can expect that
our conclusion on the uncertainty on the hadronic cross section 
due to mass effects is not going to change
significantly, since the parts of the NLO cross section that are actually
mass dependent are small.

\section{Conclusions \label{sec6}}
In this paper we  computed the virtual NLO QCD corrections in 
Higgs pair production. The double-triangle contribution was computed
exactly while  the spin-0 and
spin-2 two-loop form factors in the amplitude were computed   
via an asymptotic expansion in the top mass up to and including terms 
${\cal O} (1/\mt^8)$. Analytic results are presented for both 
contributions.
Before this  work $F_{1,2}^{2\ell}$ and $F_{1,2}^{2\Delta}$ were known only 
in the $\mt \to \infty$ limit \cite{Dawson:1998py}.
 
Our results allow a more precise evaluation of the NLO cross section for 
partonic energies up to $\sqrt{\hat{s}} \simeq 400$ GeV. This energy region 
is not  the one contributing most to the hadronic 
cross section, however, its investigation enabled us to quantify 
the difference between the NLO result obtained in the $\mt \to \infty$ limit
and the true one, where the top mass is kept finite. Although we did not
discuss the large mass evaluation of the  real contributions ${\cal R}_{gg}, \:
{\cal R}_{q \bar{q}}$ and ${\cal R}_{qg}$, their size, as estimated from
their LET values, is quite small so that even a $100 \%$ error on these
terms will not make a large difference in the hadronic cross section.  
Under the assumption that in both energy regions, 
$\sqrt{\hat{s}} \lesssim 400$ GeV and  $\sqrt{\hat{s}} \gtrsim 400$ GeV, the 
finite top-mass effects are of similar size and behavior,
we  estimated that the true NLO result is going to be smaller than the one 
obtained in the LET limit by less than $10 \%$. We remark that while
our results for $\sqrt{\hat{s}} \lesssim 400$ GeV are solid, our estimate
of the hadronic cross section, as any other based on results obtained
via a large top-mass expansion, should be understood just as a guess.

Our analysis differs in several points from previous works in the literature
\cite{Grigo:2013rya, Grigo:2015dia}. The main differences are:  i) we performed
the asymptotic expansion at the level of form factors and not of the cross
section as in refs. \cite{Grigo:2013rya, Grigo:2015dia}. ii) We did not
discuss the real contributions as instead was done in those works.
Point i) allowed us to compute the virtual NLO contribution as in 
eq.~\eqref{SCnlonoi} without making use of the factorization of the exact
LO cross section neither at the partonic level for the total cross section
as in ref.\cite{Grigo:2013rya} nor at the level of differential factorization,
i.e. before the integration over the Higgs pair invariant mass, as in 
ref.~\cite{Grigo:2015dia}. The factorization of the LO cross section is known
to work fine in single Higgs production where the exact NLO result is known
\cite{Spira:1995rr, Harlander:2005rq, Anastasiou:2006hc,Aglietti:2006tp},
however there is no proof that the same happens also in double Higgs production.
From the comparison of table \ref{tab:2} and table \ref{tab:3} in section
\ref{sec5} it seems that the differential factorization, that is expected
to lead to a better result than the other possibility since it gives rise to a 
better-behaved integrand \cite{Grigo:2015dia}, when the LET result is 
employed tends to overestimate the  result. Although a detail comparison
of our results with those of refs.~\cite{Grigo:2013rya, Grigo:2015dia} is
not possible, we notice that our results in table \ref{tab:2} and \ref{tab:3}
exhibit the  same behavior with respect to the order  of the expansion of the
soft-virtual cross section of ref.~\cite{Grigo:2015dia}.

Finally, we would like to point out that our work should be seen as one of
the  first steps towards a complete calculation of the two-loop virtual 
corrections in  Higgs pair production. 
A complete calculation of the NLO corrections,
requires to address, besides the real contributions that were studied in 
refs.~\cite{Frederix:2014hta, Maltoni:2014eza}, 
the computation of the virtual corrections in the energy
region $\sqrt{\hat{s}} \gtrsim 400 $ GeV. These corrections are very difficult 
to compute but can be attacked either via a large momentum expansion 
calculation or by numerical methods.

\subsection*{Acknowledgments}
R.G.~would like to thank Jens Hoff for clarifications concerning
refs.~\cite{Grigo:2013rya, Grigo:2015dia}. G.D.~thanks Pietro Slavich for
useful comments.
The work of P.P.G.~was partially supported  by the United States Department 
of Energy under Grant Contracts de-sc0012704. 

\clearpage
\begin{appendletterA}

\section{Two-loop form factors in the large top-mass expansion \label{app:1}}
In this appendix we provide the two-loop form factors appearing in
eq.~\eqref{f2li} expanded in the small external momenta  up to 
$\mathcal{O}(1/\mt^8)$. 
The triangle form factors read
\bea
F_{\Delta,C_{F}}^{2\ell}&=&-1+\frac{61}{270} \frac{\hat{s}}{\mt^2}+
\frac{554}{14175}\frac{\hat{s}^2}{\mt^4}+\frac{104593}{15876000}
          \frac{\hat{s}^3}{\mt^6}
                +\frac{87077}{74844000}\frac{\hat{s}^4}{\mt^8}~,
\label{fdeltaCF}\\
F_{\Delta,C_{A}}^{2\ell}&=& \frac{5}{3}+\frac{29}{1080} \frac{\hat{s}}{\mt^2}+
\frac{1}{7560} \frac{\hat{s}^2}{\mt^4}
   -\frac{29}{168000}\frac{\hat{s}^3}{\mt^6}- 
\frac{3329}{74844000} \frac{ \hat{s}^4}{\mt^8}~.
 \label{fdeltaCA} 
\eea
The spin-0 box form factors are given by
\bea
F_{\Box,C_{F}}^{2\ell}&=& 1-\frac{59 }{90} \frac{\mh^2}{\mt^2} -
    \frac{7 }{60} \frac{\hat{s}}{\mt^2} -\frac{59 }{140} \frac{\mh^4}{\mt^4}+
 \frac{551}{8100} \frac{\mh^2\,\hat{s}}{\mt^4}-
\frac{12721 }{226800} \frac{\hat{s}^2}{ \mt^4} +
\frac{251 }{5670} \frac{\pt \, \hat{s}}{\mt^4} 
 \nonumber \\ &-& 
    \frac{3821 }{22050} \frac{\mh^6}{\mt^6}+
\frac{45013}{396900} \frac{\mh^4\,\hat{s}}{\mt^6}-
\frac{229991 }{3175200} \frac{\mh^2\, \hat{s}^2}{\mt^6}+
\frac{109313 }{1587600} \frac{\mh^2\, \pt \,\hat{s}}{\mt^6}+
 \nonumber \\ &&~\frac{15829 }{15876000} \frac{\hat{s}^3}{\mt^6} 
-
  \frac{3799 }{176400} \frac{\pt \,\hat{s}^2}{\mt^6} \nonumber\\ &-&
\frac{530729 }{8731800} \frac{\mh^8}{\mt^8}+
\frac{3948398}{49116375} \frac{\mh^6\, \hat{s}}{\mt^8} -
\frac{110261363}{1571724000} \frac{\mh^4 \,\hat{s}^2}{\mt^8} 
   + \nonumber \\ && ~
   	\frac{1995199 }{37422000} \frac{\mh^4 \, \pt \,\hat{s}}{\mt^8} +
\frac{4487981 }{285768000} \frac{\mh^2 \, \hat{s}^3}{\mt^8} -
\frac{126777587}{3143448000} \frac{\mh^2\, \pt \,\hat{s}^2}{\mt^8}
- \nonumber \\ && ~
\frac{6432773 }{2095632000} \frac{\hat{s}^4}{\mt^8}	+	 
  \frac{518797 }{69854400} \frac{\pt \,\hat{s}^3}{\mt^8}-
\frac{971203 }{261954000} \frac{p_{\scriptscriptstyle{T}}^4 \,\hat{s}^2}{\mt^8}~,
\label{fboxCF} 
\\
F_{\Box,C_{A}}^{2\ell}&=& 
  	-\frac{5}{3}-\frac{139}{270} \frac{ \mh^2}{\mt^2}
  				-\frac{11}{540} \frac{\pt}{\mt^2}
   				+\frac{49}{360} \frac{\hat{s}}{\mt^2} 
 				\nonumber \\ &-& \frac{1649 }{12600}\frac{\mh^4}{\mt^4}
  				-\frac{31 }{3780} \frac{\mh^2\,\pt}{ \mt^4}
  		+ \frac{31229 }{529200} \frac{\mh^2\,\hat{s}}{\mt^4}
-
     	\frac{4499 }{105840}\frac{\pt\, \hat{s}}{\mt^4}-
\frac{451 }{88200} \frac{\hat{s}^2}{\mt^4}	
\nonumber  \\	&- &\frac{16273 }{529200} \frac{\mh^6}{\mt^6}-
\frac1{378} \frac{\mh^4 \,\pt}{\mt^6}+
\frac{516367 }{47628000} \frac{\mh^4 \,\hat{s}}{\mt^6}
 \nonumber  -
\frac{7739 }{212625} \frac{\mh^2 \,\pt \,\hat{s}}{\mt^6}+ \nonumber \\ &&~
\frac{52579 }{11907000} \frac{\mh^2\, \hat{s}^2}{\mt^6}+
\frac{103}{567000} \frac{p_{\scriptscriptstyle T}^4\,\hat{s}}{\mt^6}+
\frac{626821}{47628000} \frac{\pt\,\hat{s}^2}{\mt^6}
   -\frac{56969 }{95256000} \frac{\hat{s}^3}{\mt^6} 
  \nonumber\\  
&-&\frac{4871}{712800} \frac{\mh^8}{\mt^8}-
\frac{7}{8910} \frac{ \mh^6 \,\pt}{\mt^8}-
\frac{21577777 }{11525976000} \frac{\mh^6 \,\hat{s}}{\mt^8}-
\frac{56431033 }{2881494000 } \frac{\mh^4 \,\pt\,\hat{s}}{\mt^8}
  +  \nonumber\\ &&~
\frac{2457167}{251475840}\frac{\mh^4\,\hat{s}^2}{\mt^8}+
   	\frac{2}{12375} \frac{ \mh^2 \, p_{\scriptscriptstyle T}^4
  	\,\hat{s}}{ \mt^8}+\frac{168318277}{8644482000} \frac{\mh^2 \,\pt \,
 	\hat{s}^2}{\mt^8}-\frac{3696311}{987940800} \frac{\mh^2
	\,\hat{s}^3}{\mt^8}+
   \nonumber\\ && ~\frac{203699917}{69155856000} \frac{ p_{\scriptscriptstyle T}^4
	\,\hat{s}^2}{ \mt^8}
   -	\frac{70223597 }{17288964000} \frac{\pt\,\hat{s}^3}{\mt^8}
   	+\frac{6643339}{12573792000} \frac{\hat{s}^4}{\mt^8}\nn  \allowdisplaybreaks\\ 
&+& \log \left(\frac{\hat{s}}{\mt^2}\right)
   \left[ \frac{13 }{630} \frac{\mh^2\, \hat{s}}{\mt^4}+
\frac{13 }{420}\frac{\pt \, \hat{s}}{\mt^4}-
\frac{13 }{2520} \frac{\hat{s}^2}{\mt^4} \right.  \nn  \\
&&~~~~~~~~~~~~~~~ + \frac{2 }{105} \frac{ \mh^4\, \hat{s}}{ \mt^6}+
\frac{1}{35} \frac{\mh^2 \, \pt \,\hat{s}}{ \mt^6}-
\frac{2 }{189} \frac{ \mh^2 \, \hat{s}^2}{ \mt^6} 
  -\frac{11 }{1260} \frac{\pt \,\hat{s}^2}{ \mt^6}+
\frac{11 }{7560} \frac{\hat{s}^3}{ \mt^6} \nn \allowdisplaybreaks \\
&&~~~~~~~~~~~~~~~ + \frac{38 }{3465} \frac{\mh^6\, \hat{s}}{\mt^8}+
\frac{19}{1155} \frac{\mh^4 \,\pt \, \hat{s}}{\mt^8} 
-\frac{59 }{5670} \frac{\mh^4\,\hat{s}^2}{\mt^8} 
-\frac{797}{62370} \frac{\mh^2 \,\pt \,\hat{s}^2}{\mt^8} + \nn \\
&&~~~~~~~~~~~~~~~~~~
   	\left. \frac{83 }{24948} \frac{\mh^2 \, \hat{s}^3}{\mt^8}
 -\frac{10}{6237} \frac{ p_{\scriptscriptstyle T}^4 \,\hat{s}^2}{\mt^8}+
\frac{76  }{31185} \frac{\pt \, \hat{s}^3}{\mt^8}-
\frac{1}{2835} \frac{\hat{s}^4}{ \mt^8}\right]~, 
\label{fboxCA} 
\eea
  and the spin-2 form factors by
  \bea
  F_{2,C_{F}}^{2\ell}&=&  \frac{\pt}{\mt^2} \bigg[-\frac{131}{810}
  +\frac{338}{14175} \frac{\hat{s}}{\mt^2}-\frac{9679}{37800}\frac{ \mh^2}{\mt^2}
  \nonumber \\&-&\frac{10141}{79380}\frac{ \mh^4}{\mt^4}+
   \frac{1228043}{26460000 } \frac{\mh^2\, \hat{s}}{\mt^4}+ 
   \frac{332749}{17640000} \frac{\pt \,\hat{s}}{\mt^4}
   -\frac{1234903}{105840000}\frac{\hat{s}^2}{\mt^4}
  \nonumber \\&-&\frac{1535729}{31434480}\frac{\mh^6}{\,
   \mt^6}+
   \frac{372292}{9095625}\frac{\mh^4 \,\hat{s}}{\mt^6}
   + \frac{12986429}{561330000}\frac{ \mh^2 \, \pt \,
   \hat{s}}{\mt^6} - \nonumber \\&&~\frac{107375959}{5239080000} \frac{ \mh^2 
   \,\hat{s}^2}{\mt^6}-\frac{35525767}{5239080000}\frac{\pt
   \,\hat{s}^2}{ \mt^6}+\frac{28761377}{15717240000}\frac{\hat{s}^3}{\mt^6}
   \bigg]~, \label{f2CF} \\ 
F_{2,C_{A}}^{2\ell} &=&
   \frac{\pt}{\mt^2}  \bigg[
   \frac{308}{675}
  + \frac{1377}{9800}\frac{\mh^2}{\mt^2}+\frac{23279}{1587600}\frac{\hat{s}}{\mt^2}
   \nonumber \\ &+ &
   \frac{68777}{1905120}\frac{\mh^4}{ \mt^4}-\frac{1381031}{119070000} 
   \frac{\mh^2 \,\hat{s}}{ \mt^4}-\frac{4139287}{204120000}\frac{\pt \, \hat{s}}{\mt^4}
   +\frac{2646079}{317520000}\frac{ \hat{s}^2}{ \mt^4}
  \nonumber \\ &+& \frac{1367543}{157172400}\frac{ \mh^6}{\mt^6}
  -\frac{677103949}{57629880000}\frac{\mh^4 \, \hat{s}}{ \mt^6}
  -\frac{229643327}{14407470000}\frac{\mh^2\, \pt\,
   \hat{s}}{\mt^6} +
  \nonumber \\ && ~
    \frac{3555494423}{345779280000}\frac{\mh^2 \,
   \hat{s}^2}{\mt^6}
   + \frac{149867857}{28814940000}\frac{\pt \,
   \hat{s}^2}{ \mt^6}-\frac{129094579}{172889640000} \frac{ \hat{s}^3}{\mt^6}
  \nonumber \\ &+& \log\left(\frac{\hat{s}}{\mt^2}\right)
    \left(-\frac{121}{540}+\frac{11}{1512}
   \frac{\hat{s}}{\mt^2}-\frac{341}{3780}\frac{ \mh^2}{\mt^2} \right.\nonumber 
   \\&&~~~~~~~~~~~~~~~ -\frac{11}{378}\frac{\mh^4}{\mt^4}+
    \frac{128}{7875} \frac{\mh^2 \,\hat{s}}{\mt^4}+
    \frac{6077}{567000}\frac{\pt \, \hat{s}}{\mt^4}
  -\frac{1811}{378000}\frac{\hat{s}^2}{  \mt^4}
  \nonumber\\&&~~~~~~~~~~~~~~~ -\frac{7}{810}\frac{\mh^6}{\mt^6}+\frac{24967}{2079000}\frac{ \mh^4
   \hat{s}}{\mt^6}+\frac{118}{12375}\frac{\mh^2 \,\pt \,
   \hat{s}}{\mt^6}-\frac{5791}{891000}\frac{\mh^2 \,\hat{s}^2}{
   \mt^6}- \nonumber\\
 &&~~~~~~~~~~~~~~~~~~\left. \frac{413}{148500} \frac{\pt \, \hat{s}^2}{\mt^6}
   +\frac{7709}{12474000}\frac{ \hat{s}^3}{ \mt^6}\right) \bigg]~. 
\label{f2CA}
\eea 
\end{appendletterA}


\bibliographystyle{utphys}
\bibliography{DGG}

\end{document}